\pdfoutput=1

\documentclass[aps,prb,reprint,showpacs,superscriptaddress,floatfix]{revtex4-2}
\usepackage{graphicx}
\usepackage{graphics}
\usepackage{amsmath}
\usepackage{tabularx}
\usepackage{amssymb}
\usepackage{comment}
\usepackage{amsfonts}
\usepackage{dcolumn}
\usepackage{dsfont}
\usepackage{latexsym}
\usepackage{rotating}
\usepackage{color}
\usepackage{latexsym}
\usepackage{bbm}
\usepackage{subfigure}
\usepackage{float}
\usepackage{epsfig}
\usepackage{epsf}
\usepackage{psfrag}
\usepackage{bm}
\usepackage{amsthm}
\usepackage{eucal}
\usepackage{mathrsfs}
\usepackage{url}
\usepackage{braket}
\usepackage{array}
\usepackage[utf8]{inputenc}
\usepackage{microtype}
\usepackage{multirow}
\usepackage{comment}
\usepackage{hyperref}
\usepackage{placeins}
\hypersetup{
colorlinks=true,final=true,
        linkcolor=blue,
        citecolor=blue,
        filecolor=blue,
        urlcolor=blue,
}

\usepackage{xparse}
\usepackage{xspace}

\usepackage{ulem}

\date{\today}

\begin{document}

\title{Pressure and strain tuning of the alternating bilayer-trilayer Ruddlesden-Popper nickelate: crystal and electronic structure}
\author{Huan Wu}
\affiliation{Department of Physics, Arizona State University, Tempe, AZ 85287, USA}
\author{Yi-Feng Zhao}
\affiliation{Department of Physics, Arizona State University, Tempe, AZ 85287, USA}
\author{Antia S. Botana}
\affiliation{Department of Physics, Arizona State University, Tempe, AZ 85287, USA}

\begin{abstract}
We use first-principles calculations to investigate the crystal and electronic structure of the hybrid bilayer-trilayer Ruddlesden-Popper (RP) nickelate La$_7$Ni$_5$O$_{17}$ under hydrostatic pressure and biaxial compressive strain. By analyzing the irreducible representations of the dynamically unstable phonon modes in the high-symmetry $P4/mmm$ structure, we identify a dynamically stable lower-symmetry $C2/c$ structure containing octahedral tilts. The application of both pressure and compressive strain tends to suppress the octahedral tilts, effectively tetragonalizing the structure, in analogy with the conventional RPs. The electronic structure under hydrostatic pressure and strain has similarities, but it differs in the position of the $d_{z^2}$ bonding band from the trilayer block. This band crosses the Fermi level at a pressure of 30 GPa, but it remains below it for any level of compressive strain. This strain-induced modification mirrors the electronic structure changes observed in the conventional bilayer nickelate.

\end{abstract}

\maketitle

\section{Introduction}

The observation of superconductivity under pressure in the parent Ruddlesden–Popper (RP) $R_{n+1}$Ni$_{n}$O$_{3n+1}$ nickelates~ \cite{sun2023signatures,wang2024bulk, hou2023emergence, zhu2024superconductivity,li2023signature} opened a
new avenue for the study of high-T$_c$
superconductivity. The materials in this family are characterized by the presence of $n$-NiO$_{6}$ perovskite layers separated by rocksalt $R$O blocks along the $c$-axis \cite{greenblatt1997}. The first signatures of superconductivity were reported in the bilayer ($n=2$) RP nickelate
La$_3$Ni$_2$O$_7$~\cite{sun2023signatures, wang2024bulk, hou2023emergence} with a T$_c$ $\sim$ 80 K within a pressure range of $\sim$ 14.0 to
43.5 GPa~\cite{sun2023signatures}. 
Concomitant with the emergence of superconductivity,  a structural transition from orthorhombic $Amam$ at ambient pressure to either $Fmmm$ or tetragonal $I4/mmm$ symmetry at 14.0 GPa has been reported~\cite{sun2023signatures, wang2024structure,li327structure}. This transition is associated to a suppression of octahedral tilts, with the Ni-O-Ni bond angle changing from 168$^{\circ}$ to 180$^{\circ}$. The $n$ = 3 member La$_4$Ni$_3$O$_{10}$ also exhibits superconductivity when pressurized (albeit with lower T$_c$ $\sim$ 40 K), and a similar structural transition from monoclinic $P2_{1}/a$ at ambient pressure to tetragonal $I4/mmm$ at 13-15 GPa \cite{li2023signature,zhu2024superconductivity,li2024structural,zhang43102025,li2025trilayer}. Similar to the $n$ = 2 member, this transition is also associated with the change in the Ni-O-Ni bond angle across the apical oxygens from 166$^{\circ}$ to 180$^{\circ}$. A similar structural transition is observed in La$_3$Ni$_2$O$_7$ when grown in thin film form upon compressive strain, wherein superconductivity has also been observed \cite{ko2025signatures,bhatt2025resolving,osada2025strain,li2025angle,zhou2025ambient,liu2025superconductivity}. 
The electronic structure of both the bilayer and trilayer materials is characterized by the active role of bands of $d_{x^2-y^2}$ and $d_{z^2}$ character at the Fermi level. Many studies have suggested that the straightening of the Ni-O-Ni bond angle along the $c$-axis with pressure is related with the emergence of superconductivity due to associated enhancement in $d_{z^2}$ inter-layer coupling~\cite{luo2023bilayer, Zhang2023, Yang2023, Yang2024, Zhang2024save, sakakibara2024, christiansson2023correlated,  lechermann2023electronic, Yang2024,Zhang2024,Zhang2024save, zhang2024structural, LaBollita_2024_electronic_structure_magnetic_tendencies, Lu2024,Yang2024-orbital_selective, zhao2025electronic}.

Recently, new hybrid RP nickelate phases La$_{n+1}$Ni$_n$O$_{3n+1}$·La$_{m+1}$Ni$_m$O$_{3m+1}$ $n> m$ have been reported \cite{li2024design}. The first one to be discovered is a polymorph of La$_3$Ni$_2$O$_7$  ~\cite{1313Polymorphism202JACSmitchel,Wang_2024_Cmmm,Puphal_PhysRevLett.133.146002,arxivheptingmono-tri, huang2025superconductivitymonolayertrilayerphasela3ni2o7} whose structure is formed by alternating single-layer ($m$=1) and trilayer ($n$= 3) blocks forming a 1313 configuration. Superconductivity in this new La$_3$Ni$_2$O$_7$-1313 polymorph has also been reported under pressure with a T$_c$ as high as that of the conventional bilayer (2222) 
 counterpart ~\cite{arxivheptingmono-tri,Puphal_PhysRevLett.133.146002, huang2025superconductivitymonolayertrilayerphasela3ni2o7}. Superconductivity under pressure has also been reported in the single-layer ($m$=1) +bilayer ($n$=2) RP nickelate La$_5$Ni$_3$O$_{11}$ with a T$_c$ $\sim$ 60 K \cite{shi2025pressure}. In both of these systems, the leading space group symmetry reported at ambient pressure is $Cmmm$, in which there is no octahedral tilting at ambient pressure. There is, however, evidence in the in-lab X-ray diffraction data of weak superlattice reflections that could lead to tilted octahedra \cite{1313Polymorphism202JACSmitchel,sundaramurthy2025comparative, arxivheptingmono-tri}. The electronic structure of these hybrid phases has been scrutinized, and it exhibits some similarities with the La$_3$Ni$_2$O$_7$-2222 counterpart, while there are discrepancies about the single-layer block being in a Mott-insulating regime ~\cite{Lecherman_PRM_1313_2024,yangzhang_PRB_2023_electronic_structure_La3Ni2O7, LaBollita_PhysRevB.110.155145,arxivheptingmono-tri,Sharma20261313}. Recently, superconductivity in these two hybrid phases upon compressive strain provided by a SrLaAlO$_4$ (SLAO) substrate has been explored: while the 1212 hybrid is also superconducting in thin film form, the 1313 phase is not \cite{nie2025ambient}. 

 In order to identify strategies to emulate and surpass the superconducting properties of the existing superconducting RP nickelates, other members of the hybrid family and the effects of pressure and strain on their electronic structure can be scrutinized. As the recent results described above seem to point towards the need for a bilayer structural motif in RPs to get superconductivity (and a higher T$_c$), an obvious choice is represented by the bilayer ($m$=2) +trilayer ($n$=3) counterpart La$_7$Ni$_5$O$_{17}$-2323. This phase (not yet experimentally realized) has been scrutinized under pressure in a couple of theoretical works \cite{zhang2025magnetic,ouyang2025phase}, but no comparison with strain has been performed, and the ambient-pressure structure has not yet been carefully analyzed. The existing electronic structure calculations under pressure show the emergence of extra $d_{z^2}$ states at the Fermi level under pressure associated with the trilayer block, and  a leading $s^{\pm}$-superconducting symmetry with similar pairing strength to that obtained previously for La$_3$Ni$_2$O$_7$-2222 \cite{zhang2025magnetic}.

Here, we perform first-principles calculations to shed light on the structural and electronic structure evolution of the hybrid bilayer-trilayer RP nickelate La$_7$Ni$_5$O$_{17}$-2323 upon hydrostatic pressure and strain.  We start by investigating the structural stability of the higher-symmetry $P4/mmm$ structure at ambient pressure and explore the symmetry connections with other space groups. 
Our lattice dynamics calculations reveal that the $P4/mmm$ phase exhibits multiple dynamical instabilities. Using group theory analysis, we find that the distortions associated with the irreducible representations (irreps) of the unstable phonon branches could transform the $P4/mmm$ structure into a $C2/c$ space group, containing octahedral tilts. Under pressure and compressive strain, these octahedral tilts tend to be suppressed, tetragonalizing the structure. The electronic structure we find under pressure ($\sim$30 GPa) is similar to that reported in the literature. Upon  strain, the electronic structure differs as the bonding $d_{z^2}$ trilayer band emerging at the Fermi level upon pressure is pushed below the Fermi level upon compressive strain. This change in the electronic structure is analog to that reported in compressively strained bilayer La$_3$Ni$_2$O$_7$-2222, making it important to understand whether the related corner pockets are important to develop superconductivity.

\begin{figure}
    \centering
    \includegraphics[width = 0.5\textwidth]{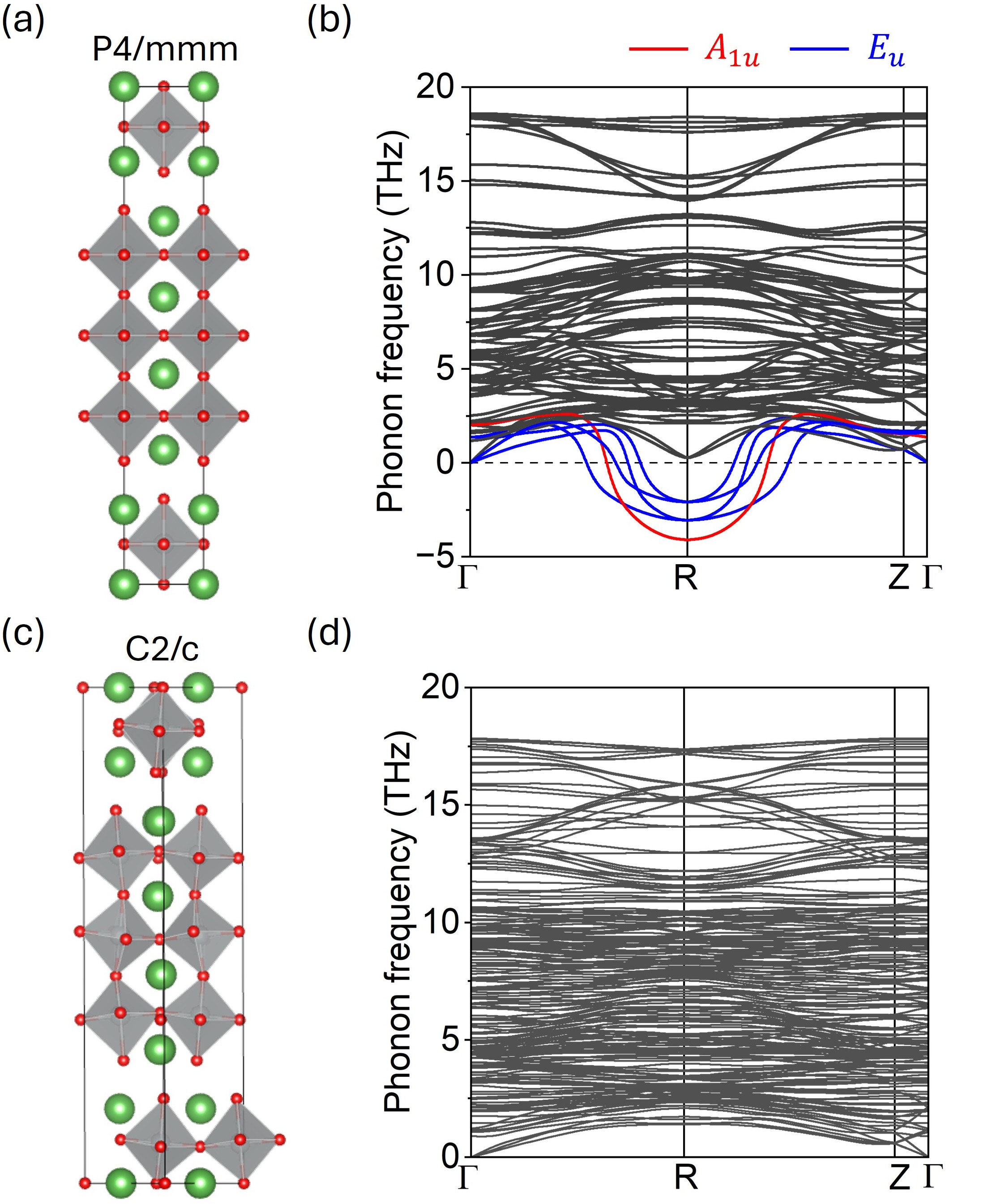}
    \caption{Crystal structure of La$_7$Ni$_5$O$_{17}$-2323 and phonon dispersion relation. The high-symmetry space group $P4/mmm$ (a), that displays no octahedral tilts, gives rise to unstable phonon modes (b). The structure obtained by applying distortions according to the irreducible representation of those modes (and relaxing) has $C2/c$ symmetry and contains octahedral tilts (c), characteristic of perovskite nickelates. The phonon dispersion of the $C2/c$ structure exhibits no imaginary frequencies (d), confirming its dynamic stability. Spheres in green, gray, and red in panels (a,c) represent the La, Ni, and O atoms, respectively. }
    \label{Fig:1}
\end{figure}

\section{Computational Methods}

The density functional theory (DFT) calculations are performed using plane-wave basis set as
implemented in Quantum ESPRESSO\cite{QE-2009, QE-2017}. The exchange-correlation is described by the
Perdew–Burke–Ernzerhof functional within the generalized gradient approximation \cite{pbe}. 
Ultrasoft pseudopotentials are employed \cite{pseudopotentials}. The kinetic energy cutoffs for the wavefunctions and for the charge density and potential are set to 80 Ry and 640 Ry, respectively. Brillouin zone integration is performed using Monkhorst-Pack $\mathbf{k}$-point grids of 12$\times$12$\times$2 and 8$\times$8$\times$2 for $P4/mmm$ and $C2/c$ structures, respectively. All structures are relaxed until the interatomic forces are less than $10^{-4}$ Ry/bohr. The Fermi energy and Fermi surface are determined from a $\mathbf{k}$-point grid of 24$\times$24$\times$2. The phonon dispersions of the $P4/mmm$ structure are calculated using ALAMODE \cite{Tadano_2014}. The interatomic force constants are derived by fitting the irreducible displacement-force set, for which the interatomic forces are calculated by DFT on a 2$\times$2$\times$1 116-atom supercell with a 6$\times$6$\times$2 Monkhorst-Pack $\mathbf{k}$-point grid.

\section{Results}
\subsection{Structural stability at ambient pressure}

\begin{figure*}
   \centering
     \includegraphics[width=0.95\linewidth]{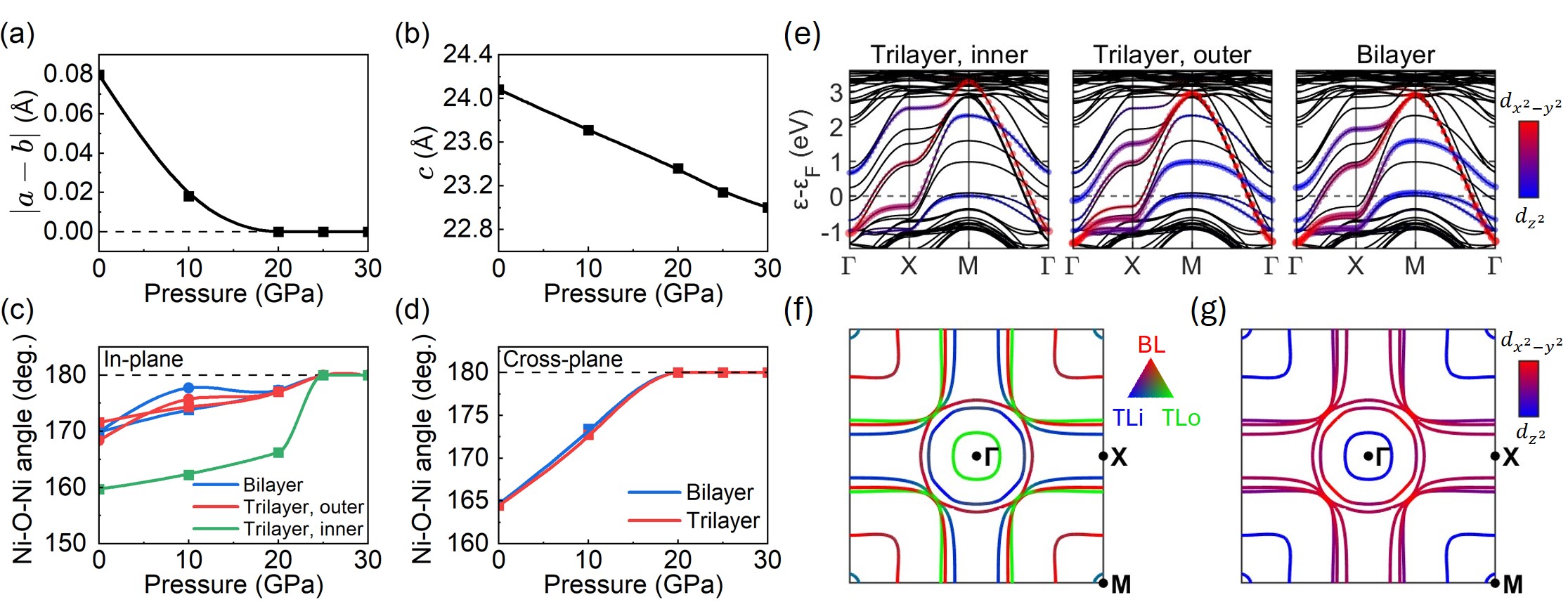}
   \caption{Evolution of the structural parameters of La$_7$Ni$_5$O$_{17}$-2323 under hydrostatic pressure and its electronic structure at 30 GPa. (a-d) Pressure-dependent evolution of (a) the absolute difference between in-plane lattice constants $|a-b|$; (b) the cross-plane lattice constant $c$; (c) the in-plane Ni-O-Ni bond angle; and (d) the cross-plane Ni-O-Ni bond angle. (e) Electronic band structure at 30 GPa, with colors representing contributions from $d_{x^2-y^2}$ (red) and $d_{z^2}$ (blue) orbitals of the Ni in the inner layer of the trilayer (left), outer layer of the trilayer (middle), and bilayer (right). (f,g) Fermi surface cutoff at the $k_z = 0$ plane, with colors representing (f) contributions from the inner layer of the trilayer (TLi), outer layer of the trilayer (TLo), and bilayer (BL) Ni sites, and (g) contributions from the $d_{x^2-y^2}$ (red) and $d_{z^2}$ (blue) orbitals of Ni.}
    \label{fig:2}
\end{figure*}

We start by analyzing the dynamical stability of the crystal structure with higher symmetry at ambient pressure ($P4/mmm$) for the bilayer-trilayer hybrid nickelate La$_7$Ni$_5$O$_{17}$, studying its phonon dispersion (see Fig.~\ref{Fig:1}(a,b)). As shown in Fig. \ref{Fig:1}(a), in tetragonal $P4/mmm$ symmetry, all Ni–O–Ni bond angles are 180$^\circ$, indicating the absence of both out-of-plane tilting and in-plane rotations of the NiO$_6$ octahedra between adjacent layers. The phonon dispersion in Fig \ref{Fig:1}(b) shows five modes exhibit imaginary frequencies at the R-point of the Brillouin zone. The lowest-energy mode corresponds to an $A_{1u}$ irreducible representation, while the remaining four modes, forming two pairs of degenerate modes at the R point,  belong to the $E_u$ irreducible representation. The distorted structures obtained by displacing atoms from their equilibrium positions according to the phonon eigenvectors of these unstable modes can be seen in Appendix \ref{appendix:A}. The modes with $A_{1u}$ symmetry correspond to an in-plane distortion of NiO$_6$ octahedra, whereas the $E_u$ ones represent tilting of the NiO$_6$ octahedra.

To identify the dynamically stable structure at ambient pressure, we generate a distorted structure on a 2$\times$2$\times$2 supercell by combining the distortions of the $A_{1u}$ mode and the two nondegenerate $E_u$ modes. After relaxation, this distorted structure converged to a $C2/c$ space group, as shown in Fig. \ref{Fig:1}(c). The phonon dispersion of the $C2/c$ structure exhibits no imaginary frequencies, confirming its dynamical stability (see Fig. \ref{Fig:1}(d)). Hence, our analysis of the unstable phonon branches in the higher-symmetry structure for La$_7$Ni$_5$O$_{17}$-2323 ($P4/mmm$) shows that it can transform into a dynamically-stable lower-symmetry $C2/c$ structure through distortions associated with the irreps of unstable modes at the R point. This hybrid structure, akin to the conventional bilayer and trilayer compounds, contains octahedral tilts with Ni-O-Ni bond angles across the apical oxygens between 160 and 170$^\circ$ (see Fig. \ref{fig:2}).

\subsection{Crystal structure and electronic structure under pressure}

\begin{figure*}
   \centering
     \includegraphics[width=\linewidth]{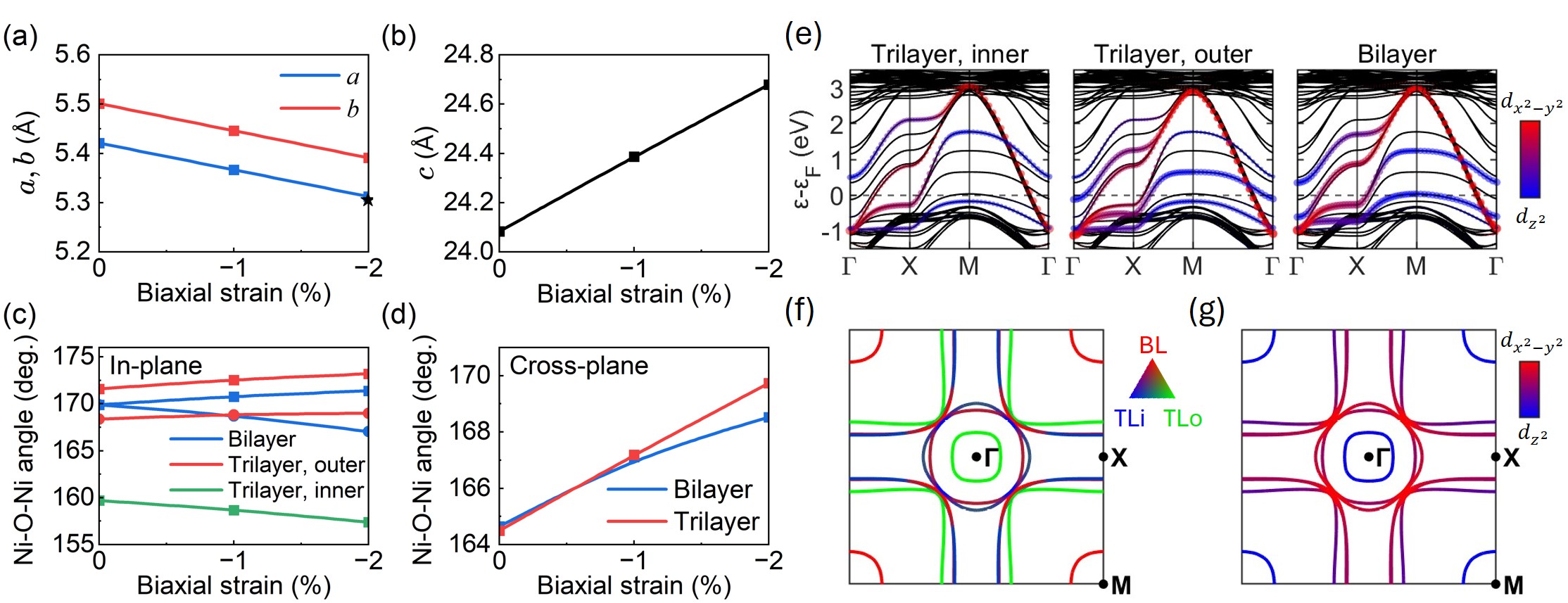}
   \caption{Evolution of the structural parameters of La$_7$Ni$_5$O$_{17}$-2323 under biaxial compressive strain and its electronic structure at a -2\% biaxial strain. (a-d) Strain-dependent evolution of (a) the in-plane lattice constants with the symmetry starting as $C2/c$; (b) the cross-plane lattice constant $c$; (c) the in-plane Ni-O-Ni bond angle; and (d) the cross-plane Ni-O-Ni bond angle. The star on the right y-axis of (a) marks the in-plane lattice constant at 15 GPa. (e) Electronic band structure at -2\% biaxial strain, with colors representing contributions from $d_{x^2-y^2}$ (red) and $d_{z^2}$ (blue) orbitals of the Ni in the inner layer of the trilayer (left), outer layer of the trilayer (middle), and bilayer (right). (f,g) Fermi surface cutoff at the $k_z = 0$ plane, with colors representing (f) contributions from the inner layer of the trilayer (TLi), outer layer of the trilayer (TLo), and bilayer (BL) Ni sites, and (g) contributions from the $d_{x^2-y^2}$ (red) and $d_{z^2}$ (blue) orbitals of Ni.}
    \label{fig:3}
\end{figure*}

After analyzing the structural stability of La$_7$Ni$_5$O$_{17}$ at ambient pressure, we now focus on its evolution with pressure. The changes under pressure for the $c$ lattice constant, the difference $\lvert$$a$-$b$$\rvert$ between the in-plane lattice constants, and the Ni-O-Ni bond angle across the apical/planar oxygens (plotted separately for the bilayer and trilayer blocks) are shown in Fig.~\ref{fig:2}(a-d). The starting (ambient pressure) structure is the $C2/c$ phase described in the previous section. The $a$ and $b$ lattice parameters become equal at $\sim$20 GPa, resulting in a tetragonalized crystal structure. Concomitantly, the Ni-O-Ni apical bond angles (cross-plane angles) straighten to 180$^{\circ}$ at 20 GPa, while for the planar angles, the transition to 180$^{\circ}$ is slower and only reached at 25 GPa. As expected, under hydrostatic pressure, all Ni-O bonds (both in-plane and cross-plane) are reduced as pressure is applied. In the context of energetics, a $P4/mmm$ structure can also be seen to become more stable at 30 GPa. These findings are consistent with the phonon calculations where, with increasing pressure, the soft phonon branches harden, leading to a dynamically stable crystal structure with $P4/mmm$ symmetry at $\sim$ 30 GPa (see Appendix~\ref{appendix:B}).

We next describe the electronic structure of La$_7$Ni$_5$O$_{17}$ under pressure (see Fig. \ref{fig:2}(e-g)), where superconductivity could emerge as pointed out by previous theory work \cite{zhang2025magnetic}. We focus on a 30 GPa pressure value , where the structure has fully transitioned to a tetragonal phase. A formal valence count renders an (average) Ni$^{2.6+}$ valence corresponding to a $d^{7.4}$ filling. With the $t_{2g}$ orbitals being fully occupied, that leaves 1.4 $e_g$ electrons per Ni to fill the $e_g$ orbitals. As a reflection of this filling, the electronic structure of La$_7$Ni$_5$O$_{17}$ shows bands of both $d_{z^2}$ and $d_{x^2-y^2}$ character crossing the Fermi level. These states are highly hybridized with O-$p$ states. Like in the isolated/conventional trilayer and bilayer compounds, the $d_{z^2}$ orbitals are split into a molecular orbital combination of bonding-antibonding states (bilayer block) and of bonding-nonbonding-antibonding states (trilayer block). The bonding $d_{z^2}$ orbital from the bilayer clearly crosses the Fermi level  at the M point of the Brillouin zone, giving rise to a corner pocket in the Fermi surface. The bonding $d_{z^2}$ orbital from the trilayer barely crosses at 30 GPa, giving rise to a very small hole pocket also at the zone corner. As described before \cite{zhang2025magnetic,ouyang2025phase} the emergence of this trilayer band at the Fermi level is the most important change with pressure (in addition to the expected bandwidth increase). While the size of this pocket at 30 GPa is smaller in our calculations, in prior work using random phase approximation (RPA) calculations, its emergence was deemed to be important for superconductivity \cite{zhang2025magnetic}. The gap distribution found in this work is very similar to that of the isolated trilayer -- but with an enhanced pairing strength. We note, however, that this pocket is fragile and the introduction of an on-site Coulomb repulsion $U$ (as well as a lower pressure) can push the bonding trilayer band below the Fermi level (see Appendices \ref{appendix:C} and \ref{appendix:D}). The $d_{z^2}$ bands higher in energy (above the Fermi level) are the nonbonding one from the trilayer (with only contributions from the outer Ni), the antibonding one from the bilayer, and finally the antibonding band from the trilayer Nis.  Ultimately, the low-energy physics in the DFT nonmagnetic bands seems to be a combination of both of the constituent structural blocks (with some small band shifts).

The derived hoppings at 30 GPa further reinforce this picture of superposition of trilayer and bilayer physics in La$_7$Ni$_5$O$_{17}$ (see Table \ref{hoppings}). The dominant hoppings obtained for the ``isolated'' conventional bilayer La$_3$Ni$_2$O$_7$ and trilayer La$_4$Ni$_3$O$_{10}$ are almost identical to those derived for the 2323 material at 30 GPa-- and these values significantly increase from their ambient pressure counterparts. Such dominant hoppings are $t_{\perp}^z$ that defines the overlap between the $d_{z^2}$ orbitals via the O-$p_z$ orbitals ($\sim$ 0.6 eV at 30 GPa), the nearest-neighbor hopping for the $d_{x^2-y^2}$ orbitals $t_{\parallel}^x$ $\sim$ 0.5 eV under pressure, and the hybridization in-plane between $d_{x^2-y^2}$ and $d_{z^2}$ orbitals $t_{\parallel}^{xz}$ $\sim$ 0.2 eV also at 30 GPa. 

 \begin{table}
    \centering
    \caption{Hopping parameters (in eV) for La$_7$Ni$_5$O$_{17}$, bilayer La$_3$Ni$_2$O$_7$, and trilayer La$_4$Ni$_3$O$_{10}$ under ambient and high pressure (15 and 30 GPa), as well as upon a -2\% compressive strain.}
    \resizebox{8.5cm}{!}{
    \renewcommand{\arraystretch}{1.2}
    \begin{tabular}{cc|c|cccc|cc|cc}
    \hline
    \hline
         & & \multirow{2}{*}{\textbf{Ni}} 
         & \multicolumn{4}{c|}{\textbf{hybrid bilayer+trilayer}} & \multicolumn{2}{c|}{\textbf{bilayer}} & \multicolumn{2}{c}{\textbf{trilayer}} \\
         & & & 0 GPa 
        & 15 GPa & 30 GPa & -2\% & 30 GPa & -2\% & 30 GPa & -2\% \\
        \hline
        & \multirow{2}{*}{$t_{\perp}^z$}
        &TL& -0.595 & -0.639 & -0.677 & -0.606 & & & -0.676 & -0.605 \\
        & &BL& -0.581 & -0.627 & -0.664 & -0.591 & -0.665 & -0.596 & &  \\
        \hline
        & \multirow{3}{*}{$t_{\parallel}^x$}
        &TLi& -0.441 & -0.479 & -0.510 & -0.479 & & & -0.512 & -0.480 \\
        & &TLo& -0.442 & -0.478 & -0.507 & -0.476 & & & -0.507 & -0.474 \\
        & &BL& -0.438 & -0.475 & -0.507 & -0.470 & -0.507 & -0.475& &  \\
        \hline
        & \multirow{3}{*}{$t_{\parallel}^{xz}$}
        &TLi& 0.240 & 0.261 & 0.277 & 0.244 & & & 0.280 & 0.245 \\
        & &TLo& 0.229 & 0.250 & 0.265 & 0.225 & & & 0.269 & 0.227 \\
        & &BL& 0.223 & 0.242 & 0.258 & 0.212 & 0.255 & 0.208& &  \\
    \hline
    \end{tabular}
    }
    \label{hoppings}
\end{table}

\subsection{Crystal structure and electronic structure under strain}

We continue by looking at the structural data for La$_7$Ni$_5$O$_{17}$ under compressive strain (see Fig. \ref{fig:3}(a-d)). As mentioned above, superconductivity has been achieved in bilayer La$_3$Ni$_2$O$_{7}$, as well as in the single-layer bilayer hybrid La$_5$Ni$_3$O$_{11}$ thin films upon a -2\% compressive strain by growing thin films on SLAO \cite{ko2025signatures,bhatt2025resolving,osada2025strain,li2025angle,zhou2025ambient,liu2025superconductivity,nie2025ambient}. The relationship between the emergence of superconductivity and the structural changes upon strain has been carefully scrutinized at least for the bilayer material suggesting that the straightening of the
Ni-O-Ni bond angles is related to the emergence of superconductivity in strained thin films \cite{bhatt2025resolving}. As shown in Fig. \ref{fig:3}(c,d), we
find that compressive strain in La$_7$Ni$_5$O$_{17}$ also straightens the apical Ni-O-Ni bond angle. However, this trend is not systematic for the planar angles, as the inner layer of the trilayer tends to buckle slightly upon compressive strain, and so does one of the two inequivalent angles in the bilayer block. We note that the in-plane structural parameters for a -2\% compressive strain approximately match the lattice constants obtained for a pressure of $\sim$ 15 GPa, as shown in Fig. \ref{fig:3}(a) (a pressure value that is also not fully in the tetragonal regime).

In Fig. \ref{fig:3}(e-g), the band structures along high-symmetry directions and the corresponding Fermi surfaces are shown for La$_7$Ni$_5$O$_{17}$ for a -2\% compressive strain. In order to make the comparison with pressure clearer, both of these cases are shown in $P4/mmm$ symmetry as this is the space group attained under pressure, as explained above. Once again,  the band structure near the Fermi level is characterized by Ni-$e_g$ ($d_{z^2}$ and $d_{x^2-y^2}$)  states, indicating that these two Ni-$d$ orbitals still play the dominant role in the low-energy electronic structure of this material under strain. While the electronic structure is similar to that described under pressure, the $d_{z^2}$ bonding bands are pushed down to lower energies so that at a -2\% compressive strain, only the bilayer bonding band crosses it. The same effect has been shown in the isolated bilayer and trilayer RP phases under compressive strain from DFT calculations \cite{zhao2025electronic}. The electronic structure at this strain mimics better that obtained at 15 GPa (with some differences in band splittings), as shown in Appendix \ref{appendix:D} -- as mentioned above, the in-plane lattice constants also match those at 15 GPa, as shown in Fig. \ref{fig:3}(a). The same applies to the hopping values obtained for a -2\% strain that are lower than those obtained at 30 GPa and more similar instead to those derived at 15 GPa, as shown in Table \ref{hoppings}.  Significantly lower values are obtained for $t_{\perp}^z$, as expected, given the elongation of the $c$ lattice constant upon compressive strain. The planar hoppings can instead be closely matched to the corresponding 15 GPa pressure. In the context of our electronic structure description, if superconductivity can be obtained upon strain in the 2323 RP nickelate polymorph La$_7$Ni$_5$O$_{17}$, it will be interesting to understand what role (if any) the extra corner pocket from the trilayer block might play for superconductivity when contrasting to pressure and also what the contributions of bilayer and trilayer blocks need to be in order for superconductivity to emerge. 

\textit{Note added}. During the completion of this work, evidence for superconductivity in La$_7$Ni$_5$O$_{17}$-2323 was demonstrated by the authors of Ref. \cite{nie2025ambient}, as presented in the Frontiers of Unconventional Superconductivity Virtual Workshop (work unpublished).

\vspace{0.1in}

\section{Conclusions}
In summary, we have systematically studied the effects of pressure and biaxial compressive strain in the hybrid bilayer-trilayer (2323) RP nickelate  La$_7$Ni$_5$O$_{17}$ via first-principles calculations. We have found that the structure of this hybrid nickelate at ambient pressure contains octahedral tilts, akin to the conventional trilayer and bilayer RP nickelates. Upon pressure, the structure becomes tetragonal, with the tilts of oxygen octahedra being completely suppressed at $\sim$ 25 GPa. Under compressive strain, while the out-of-plane tilts are suppressed, the planar ones remain (at least up to the highest level of strain we have studied of -2\%).  The electronic structure of La$_7$Ni$_5$O$_{17}$
resembles a superposition of the conventional bilayer and trilayer RP components with some band shifts. At high pressures, an extra corner pocket of pure $d_{z^2}$ character from the trilayer block emerges, but it is fragile as it can easily be shifted down to lower energies with a small $U$. This pocket is absent under compressive strain. If this hybrid system can be synthesized in bulk or thin film form, it would then be interesting to understand what role (if any) this extra corner pocket from the trilayer block can play for superconductivity.

\section*{acknowledgements} 
We would like to thank Harry LaBollita for fruitful discussions. We acknowledge NSF grant No. DMR-2045826 and the ASU research computing center for HPC resources.

\bibliography{ref.bib}

\onecolumngrid

\newpage
\appendix

\section{Structural distortions from imaginary phonon modes}\label{appendix:A}

\begin{figure*}[!htbp]
   \centering
     \includegraphics[width=0.6\linewidth]{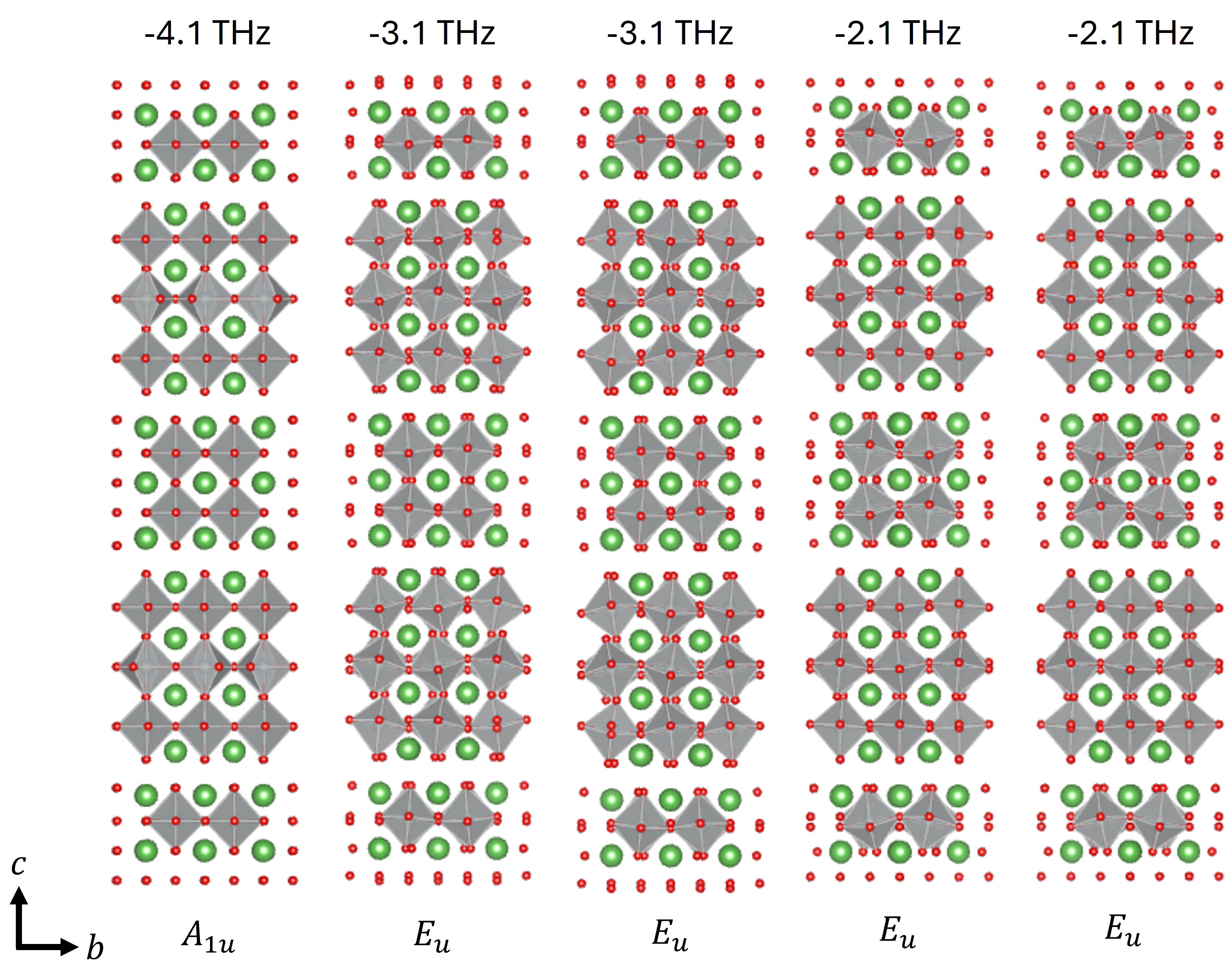}
   \caption{Distortions based on the eigenvectors of the phonon modes with imaginary frequencies (labeled on top) at ambient pressure for La$_7$Ni$_5$O$_{17}$-2323 in $P4/mmm$ symmetry. The corresponding irreducible representations are labeled at the bottom.}
    \label{fig:A1}
\end{figure*}

Figure. \ref{fig:A1} demonstrates the lattice distortions associated with the five dynamically unstable modes of the $P4/mmm$ structure of La$_7$Ni$_5$O$_{17}$-2323  at ambient pressure. The most unstable mode, at -4.1 THz, has $A_{1u}$ symmetry and is characterized by in-plane rotations within the inner layer of the trilayer, with NiO$_6$ octahedra rotating in opposite directions both within each layer and between vertically aligned octahedra in adjacent layers. The other four imaginary-frequency modes form two doubly degenerate $E_u$ pairs at -3.1 THz and -2.1 THz. These modes involve antiphase octahedral tilts, in which adjacent octahedra tilt in opposite directions within the corner-sharing octahedral framework. Within each degenerate pair, the two modes correspond to symmetry-related octahedral tilting patterns. The -3.1 THz modes exhibit stronger distortions in the trilayer, whereas the -2.1 THz modes show larger distortions in the bilayer. The bilayer-trilayer tilting correlation is reversed between the -3.1 and -2.1 THz modes.

\section{Evolution of the phonon dispersion in $P4/mmm$ symmetry with pressure}\label{appendix:B}

\begin{figure*}[!htbp]
   \centering
     \includegraphics[width=0.8\linewidth]{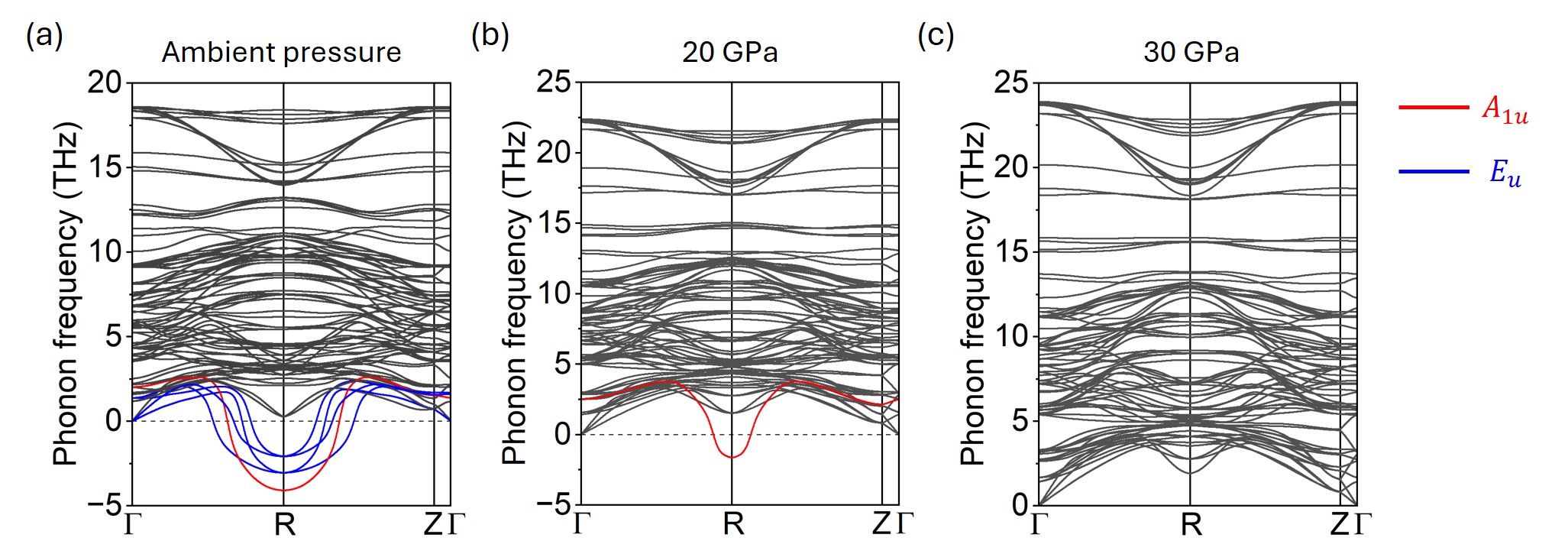}
   \caption{Phonon dispersion of La$_7$Ni$_5$O$_{17}$-2323 in $P4/mmm$ symmetry  under (a) ambient pressure, (b) 20 GPa, and (c) 30 GPa.}
    \label{fig:A2}
\end{figure*}

Figure. \ref{fig:A2} demonstrates the evolution of the phonon dispersion of La$_7$Ni$_5$O$_{17}$-2323 in $P4/mmm$ symmetry with pressure. At ambient pressure, one $A_{1u}$ mode and two doubly degenerated $E_u$ modes exhibit imaginary frequencies, as explained in the main text. At 20 GPa only the the $A_{1u}$ mode associated with in-plane rotations remains imaginary. Consistently, the structure with lowest electronic energy at 20 GPa shows in-plane octahedra rotations without cross-plane tilting. At 30 GPa, the phonon dispersion demonstrates no imaginary frequencies, confirming the dynamical stability of the $P4/mmm$ phase at 30 GPa.

\FloatBarrier
\section{Electronic structures considering DFT+$U$}\label{appendix:C}

\begin{figure*}[!htbp]
   \centering
     \includegraphics[width=0.5\linewidth]{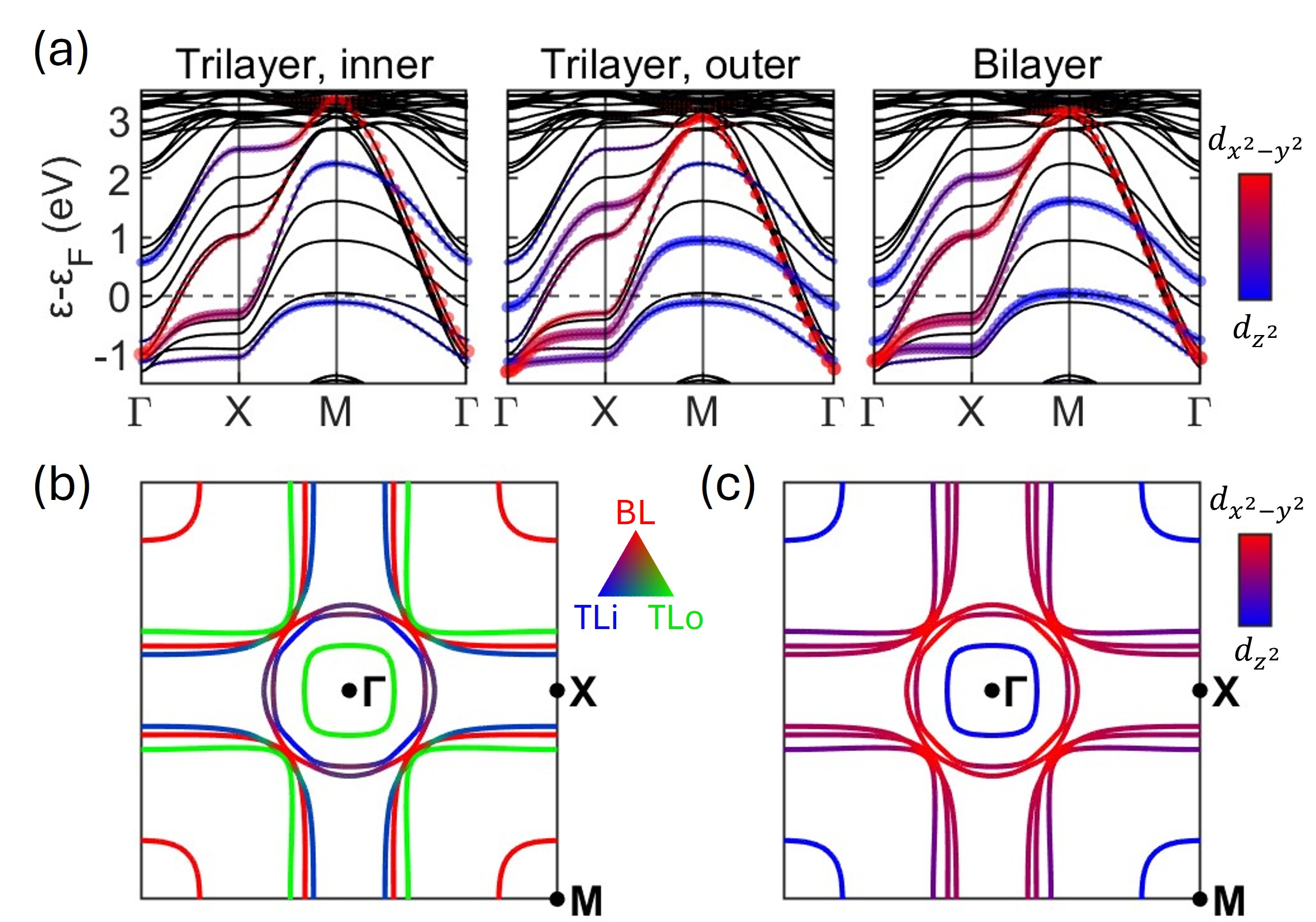}
   \caption{Electronic structure of La$_7$Ni$_5$O$_{17}$-2323 in $P4/mmm$ symmetry under 30 GPa with $U$ = 4.7 eV. (a) Electronic band structure. (b,c) Fermi surface cutoff at the $k_z = 0$ plane. The color scale represents contributions from different Ni orbitals or sites.}
    \label{fig:A5}
\end{figure*}

\begin{figure*}[!htbp]
   \centering
     \includegraphics[width=0.5\linewidth]{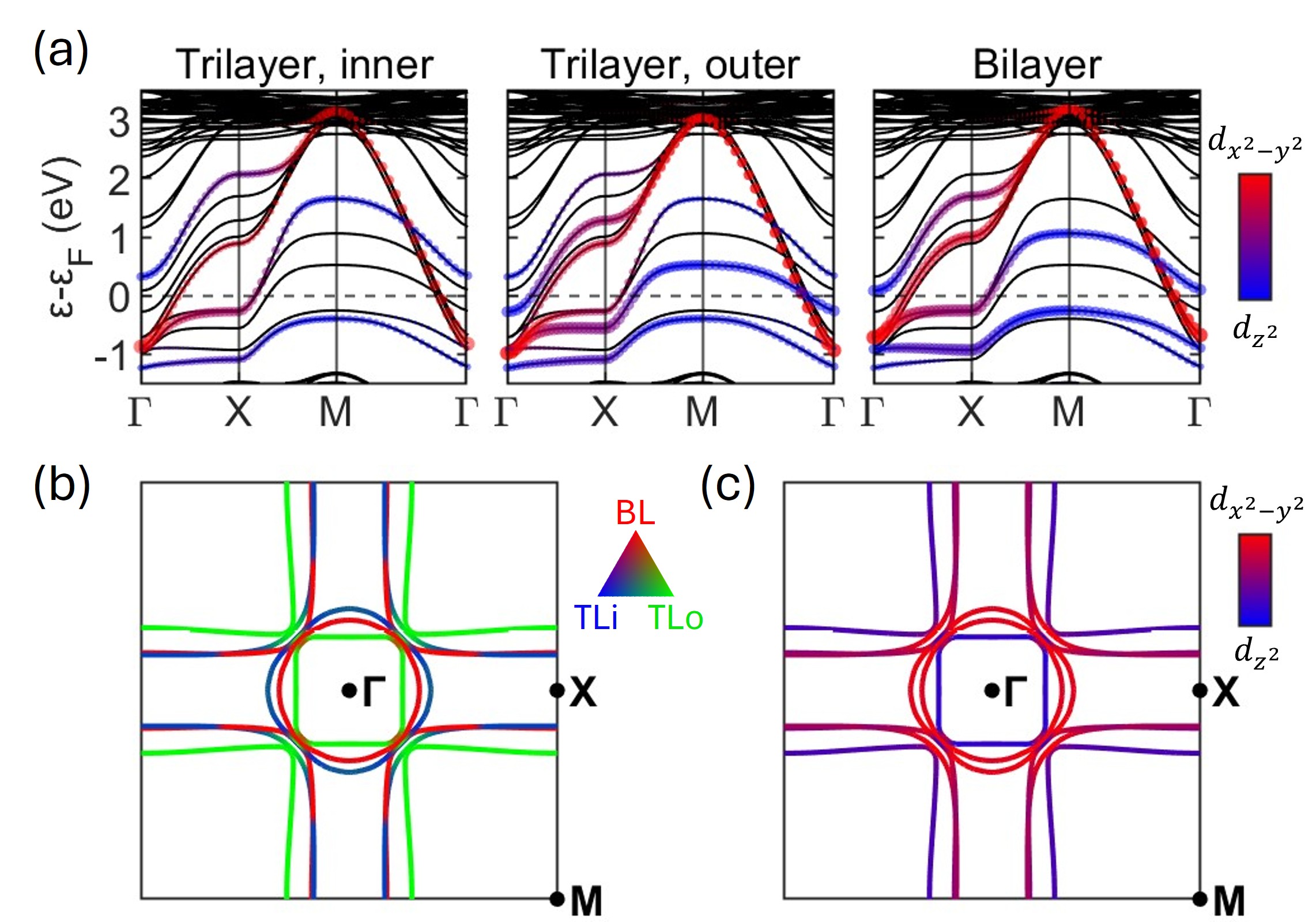}
   \caption{Electronic structure of La$_7$Ni$_5$O$_{17}$-2323 in $P4/mmm$ symmetry under a -2\% biaxial compressive strain with $U$ = 4.7 eV. (a) Electronic band structure. (b,c) Fermi surface cutoff at the $k_z = 0$ plane. The color scale represents contributions from different Ni orbitals or sites.}
    \label{fig:A6}
\end{figure*}

The electronic  structure for La$_7$Ni$_5$O$_{17}$-2323 in $P4/mmm$ symmetry with a $U$ = 4.7 eV under 30 GPa, and -2\% biaxial strain is shown in Fig. \ref{fig:A5}, and Fig. \ref{fig:A6}, respectively. The DFT+$U$ calculations were performed using the Dudarev formulation \cite{Dudarev}. A Hubbard $U$ of 4.7 eV was applied to Ni-$3d$ manifold. Considering DFT+$U$ shifts the $d_{z^2}$ bands to lower energy.


\FloatBarrier
\section{Electronic structure at 15 GPa}\label{appendix:D}

\begin{figure*}[!htbp]
   \centering
     \includegraphics[width=0.5\linewidth]{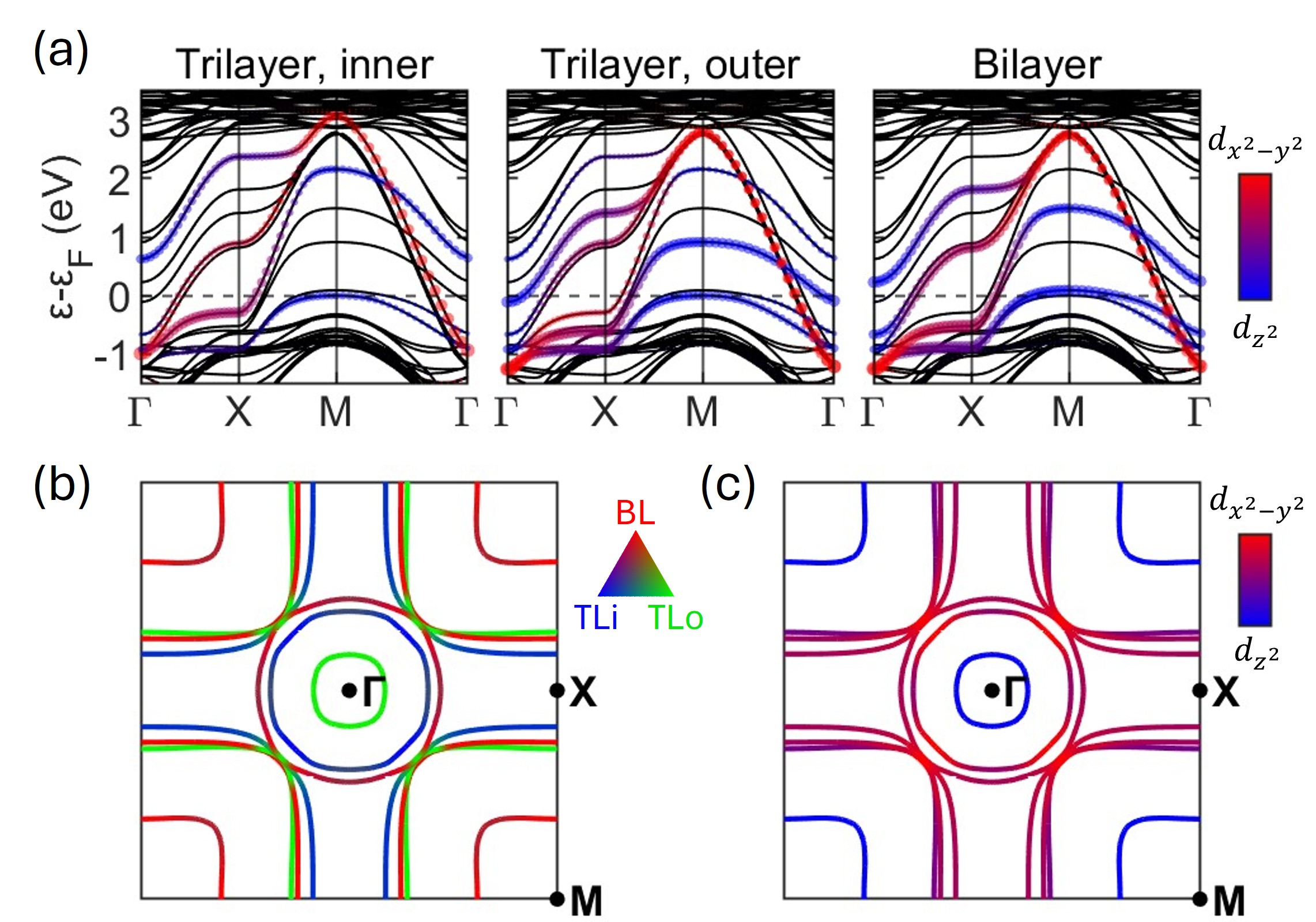}
   \caption{Electronic structure of La$_7$Ni$_5$O$_{17}$-2323 in $P4/mmm$ symmetry under 15 GPa. (a) Electronic band structure. (b,c) Fermi surface cutoff at the $k_z = 0$ plane. The color scale represents contributions from different Ni orbitals or sites.}
    \label{fig:A3}
\end{figure*}

Figure \ref{fig:A3} shows the electronic structure of La$_7$Ni$_5$O$_{17}$-2323 in $P4/mmm$ symmetry at 15 GPa, which mimics closely the lattice constants obtained under a -2\% compressive biaxial strain.

\FloatBarrier

\end{document}